\documentclass[english]{article}
\usepackage[T1]{fontenc}
\usepackage[latin9]{inputenc}
\usepackage{color}
\usepackage{textcomp}
\usepackage{amstext}
\usepackage{amssymb}
\usepackage{esint}

\makeatletter

\newcommand{\lyxmathsym}[1]{\ifmmode\begingroup\def\b@ld{bold}
  \text{\ifx\math@version\b@ld\bfseries\fi#1}\endgroup\else#1\fi}

\makeatother

\usepackage{babel}
\begin{document}

\title{\textbf{Time dependent Schrödinger equation for black hole evaporation:
no information loss}}

\author{\textbf{Christian Corda}}

\maketitle
\begin{center}
Dipartimento di Fisica e Chimica, Scuola Superiore Internazionale
di Studi Universitari e Ricerca ``Santa Rita'', Centro di Scienze
Naturali, Via di Galceti, 74, 59100 Prato, Italy
\par\end{center}

\begin{center}
Institute for Theoretical Physics and Advanced Mathematics (IFM) Einstein-Galilei,
Via Santa Gonda 14, 59100 Prato, Italy
\par\end{center}

\begin{center}
International Institute for Applicable Mathematics \& Information
Sciences (IIAMIS),  B.M. Birla Science Centre, Adarsh Nagar, Hyderabad
- 500 463, India 
\par\end{center}

\begin{center}
\textit{E-mail address:} \textcolor{blue}{cordac.galilei@gmail.com} 
\par\end{center}
\begin{abstract}
In 1976 S. Hawking claimed that ``\emph{Because part of the information
about the state of the system is lost down the hole, the final situation
is represented by a density matrix rather than a pure quantum state}''%
\footnote{Verbatim from ref. 2%
}. This was the starting point of the popular \textquotedblleft{}black
hole (BH) information paradox\textquotedblright{}.

In a series of papers, together with collaborators, we naturally interpreted
BH quasi-normal modes (QNMs) in terms of quantum levels discussing
a model of excited BH somewhat similar to the historical semi-classical
Bohr model of the structure of a hydrogen atom. Here we explicitly
write down, for the same model, a \emph{time dependent Schrödinger
equation }for the system composed by Hawking radiation and BH QNMs.
The physical state and the correspondent\emph{ wave function }are
written in terms of an \emph{unitary} evolution matrix instead of
a density matrix. Thus, the final state results to be a \emph{pure}
quantum state instead of a mixed one. Hence, Hawking's claim is falsified
because BHs result to be well defined quantum mechanical systems,
having ordered, discrete quantum spectra, which respect 't Hooft's
assumption that Schröedinger equations can be used universally for
all dynamics in the universe. As a consequence, information comes
out in BH evaporation in terms of pure states in an unitary time dependent
evolution. 

In Section 4 of this paper we show that the present approach permits
also to solve the entanglement problem connected with the information
paradox.
\end{abstract}
\textbf{\emph{To the memory of the latter IFM Secretary Franco Pettini.}}

\section{Introduction: the black hole information paradox}

One of the most famous and intriguing scientific controversies in
the whole history of Science is the so called \textquotedblleft{}BH
information paradox\textquotedblright{}. In classical gravity, a BH
is the definitive prison. Nothing can escape from it. Thus, when matter
disappears into a BH, the information encoded is considered as preserved
inside it, although inaccessible to outside observers. The situation
radically changed when Hawking discovered that quantum effects cause
the BH to emit radiation \cite{key-1}. A further analysis, again
by Hawking \cite{key-2}, has shown that the detailed form of the
radiation emitted by a BH should be thermal and independent of the
structure and composition of matter that collapsed to form the BH.
Hence, the radiation state is considered a completely mixed one which
cannot carry information about how the BH is formed. 

After Hawking's original claim, enormous time and effort were and
are currently devoted to solve the paradox. Notice that consequences
of the BH information puzzle are not trivial. If information was lost
in BHs, pure quantum states arising from collapsed matter would decay
into mixed states arising from BH evaporation and quantum gravity
wouldn\textquoteright{}t be unitary \cite{key-3}! Various scientists
worked and currently work on this issue. Some of them remained convinced
that quantum information should be destroyed in BH evaporation. Other
ones claimed that Hawking's original statement was wrong and information
should be, instead, preserved. Susskind wrote a pretty and popular
science book on details of the so called ``Black Hole War'' \cite{key-4}.
In fact, the paradox was introduced into physics folklore \cite{key-4,key-5}.
Hawking made two famous bets, one, with Thorne like co-signer, with
Preskill, another with Page, that BH does destroy information \cite{key-4}.
After almost 30 years, Hawking reversed his opinion and agreed that
information would probably be recovered \cite{key-3}. Historical
notes on the paradox's controversy and on various attempts to solve
it can be found in \cite{key-3}-\cite{key-7}. Recently, Hawking
changed again his opinion by verbatim claiming that ``there is effective
information loss'' \cite{key-36}.

A key point, not only in the framework of the BH information paradox,
but in the whole BH quantum physics, is that analysing Hawking radiation
as tunnelling, Parikh and Wilczek showed that the radiation spectrum
cannot be strictly thermal \cite{key-8,key-9}, differently from Hawking's
original computations \cite{key-1,key-2}. The energy conservation
enables the BH to contract during the process of radiation \cite{key-8,key-9}.
Thus, the horizon recedes from its original radius to a new, smaller
radius \cite{key-8,key-9}. As a consequence, BHs cannot strictly
emit thermally \cite{key-8,key-9}. This is consistent with unitarity
\cite{key-8} and has profound implications for the BH information
paradox. In fact, Parikh \cite{key-8} correctly stresses that arguments
that information is lost during BH evaporation rely in part on the
assumption of strict thermal behavior of the spectrum \cite{key-2}.
Assuming the non-thermal spectrum of Parikh and Wilczek, Zhang, Cai,
Zhan and You recently found the existence of correlations among Hawking
radiations which are elegantly described as hidden messengers in BH
evaporation \cite{key-10,key-11,key-12}. Thus, they claimed to have
recovered the information loss in Hawking radiation and to have solved
the paradox \cite{key-10,key-11,key-12}. This issue generated a controversy
with Mathur, who claimed that, instead, they failed to solve the paradox
\cite{key-13}. Mathur thinks that the foundation of the information
problem is the growing entanglement entropy between the inside and
outside of the BH \cite{key-7,key-13}. He also claims that only string
theory, in the framework of the so called ``fuzzball'', can ultimately
solve the paradox. The subsequent strong rebuttal by Zhang, Cai, Zhan
and You \cite{key-14} claims that, instead, Mathur's argument on
the growing entanglement entropy is wrong and the correlations that
they found are sufficient to resolve the paradox. This controversy
has an important scientific value, as the contenders received the
First Award \cite{key-15} and Third Award \cite{key-16} respectively
in the important Gravity Research Foundation Essay Competition 2013
for their works on the information paradox. In this work, we discuss
our proposal to solve the information puzzle. Following 't Hooft \cite{key-6},
we think that the foundation of the BH information problem is that
BHs look to do not obey Schröedinger equations, which would allow
pure states to evolve only into pure states. They look indeed to obey
probabilistic equations of motion that are not purely quantum mechanical
\cite{key-6}. We will show that, instead, a time dependent Schröedinger
equation allowing pure states to evolve only into pure states can
be found in the correspondence between Hawking radiation and BH QNMs
that we recently discussed in a series of papers \cite{key-17,key-18,key-19},
also together with collaborators \cite{key-20,key-38,key-39}. In
those papers, we naturally interpreted BH QNMs in terms of quantum
levels discussing a model of excited BH somewhat similar to the historical
semi-classical Bohr model of the structure of a hydrogen atom \cite{key-31,key-32}.
In Section 4 of this paper, we show that the present approach permits
also to solve the entanglement problem, discussed in \cite{key-13},
connected with the information puzzle. 

We also stress that consistence between the analysis in the present
paper and a recent approach to solve the BH information paradox \cite{key-10,key-11,key-12,key-14}
has been recently highlighted in \cite{key-40}.

\section{Quasi-normal modes in non-thermal approximation}

BH QNMs are frequencies of radial spin-$j$ perturbations which obey
a time independent Schröedinger-like equation, see \cite{key-17,key-18,key-19,key-21}.
They are the BH modes of energy dissipation which frequency is allowed
to be complex \cite{key-21}. The intriguing idea to model the quantum
BH in terms of BH QNMs arises from a remarkable paper by York \cite{key-22}.
For large values of the quantum ``overtone'' number $n$, where
$n=1,2,...$, QNMs become independent of both the spin and the angular
momentum quantum numbers \cite{key-17,key-18,key-19,key-21,key-23,key-24},
in perfect agreement with \emph{Bohr's Correspondence Principle} \cite{key-33},
which states that \textquotedblleft{}transition frequencies at large
quantum numbers should equal classical oscillation frequencies\textquotedblright{}.
In other words, Bohr's Correspondence Principle enables an accurate
semi-classical analysis for large values of the principal quantum
number $n,$ i.e, for excited BHs. By using that principle, Hod has
shown that QNMs release information about the area quantization as
QNMs are associated to absorption of particles \cite{key-23}. Hod's
work was refined by Maggiore \cite{key-24} who solved some important
problems. On the other hand, as QNMs are \emph{countable} frequencies,
ideas on the \emph{continuous} character of Hawking radiation did
not agree with attempts to interpret QNMs in terms of emitted quanta,
preventing to associate QNMs modes to Hawking radiation \cite{key-21}.
Recently, Zhang, Cai, Zhan and You \cite{key-10,key-11,key-12,key-14,key-15}
and ourselves and collaborators \cite{key-17,key-18,key-19,key-20}
observed that the non-thermal spectrum of Parikh and Wilczek also
implies the countable character of subsequent emissions of Hawking
quanta. This issue enables a natural correspondence between QNMs and
Hawking radiation, permitting to interpret QNMs also in terms of emitted
energies \cite{key-17,key-18,key-19,key-20}. In fact, QNMs represent
the BH reaction to small, discrete perturbations in terms of damped
oscillations \cite{key-17,key-18,key-19,key-20,key-21,key-22,key-23,key-24}.
The capture of a particle which causes an increase in the horizon
area is a type of discrete perturbation \cite{key-21,key-22,key-23,key-24}.
Then, it is very natural to assume that the emission of a particle
which causes a decrease in the horizon area is also a perturbation
which generates a reaction in terms of countable QNMs as it is a discrete
instead of continuous process \cite{key-17,key-18,key-19,key-20}.
On the other hand, the correspondence between emitted radiation and
proper oscillation of the emitting body is a fundamental behavior
of every radiation process in Science. Based on such a natural correspondence
between Hawking radiation and BH QNMs, one can consider QNMs in terms
of quantum levels also for emitted energies \cite{key-17,key-18,key-19,key-20}.
This important point is in agreement with the general idea that BHs
can be considered in terms of highly excited states in an underlying
quantum gravity theory \cite{key-17,key-18,key-19,key-20}. 

\noindent Working with $G=c=k_{B}=\hbar=\frac{1}{4\pi\epsilon_{0}}=1$
(Planck units), in strictly thermal approximation the probability
of emission of Hawking quanta is \cite{key-1,key-8,key-9} 
\begin{equation}
\Gamma\sim\exp(-\frac{\omega}{T_{H}}),\label{eq: hawking probability}
\end{equation}

\noindent where $\omega$ is the energy-frequency of the emitted particle
and $T_{H}\equiv\frac{1}{8\pi M}$ is the Hawking temperature. By
taking into account the energy conservation, i.e. the BH contraction
which enables a varying geometry, one gets the fundamental correction
of Parikh and Wilczek \cite{key-8,key-9} 
\begin{equation}
\Gamma\sim\exp[-\frac{\omega}{T_{H}}(1-\frac{\omega}{2M})],\label{eq: Parikh Correction}
\end{equation}
where the additional term $\frac{\omega}{2M}\:$ is present. We have
recently finalized the Parikh and Wilczek tunnelling picture showing
that the probability of emission (\ref{eq: Parikh Correction}) is
indeed associated to the two distributions \cite{key-25} 
\begin{equation}
<n>_{boson}=\frac{1}{\exp\left[4\pi\left(2M-\omega\right)\omega\right]-1},\;\;<n>_{fermion}=\frac{1}{\exp\left[4\pi\left(2M-\omega\right)\omega\right]+1},\label{eq: final distributions}
\end{equation}
for bosons and fermions respectively, which are \emph{non}-strictly
thermal. It is important to stress that, as it is always $\omega\leq M$
because the BH cannot emit more energy than its total mass, we have
neither negative number of particles nor divergences for finite values
of $\omega>0$ in eq. (\ref{eq: final distributions}) \cite{key-42}.

\noindent By introducing the \emph{effective temperature }\cite{key-17,key-18,key-19,key-20,key-25}
\begin{equation}
T_{E}(\omega)\equiv\frac{2M}{2M-\omega}T_{H}=\frac{1}{4\pi(2M-\omega)},\label{eq: Corda Temperature}
\end{equation}
one rewrites eq. (\ref{eq: Corda Temperature}) in a Boltzmann-like
form similar to eq. (\ref{eq: hawking probability}) 
\begin{equation}
\Gamma\sim\exp[-\beta_{E}(\omega)\omega]=\exp(-\frac{\omega}{T_{E}(\omega)}),\label{eq: Corda Probability}
\end{equation}

\noindent where $\exp[-\beta_{E}(\omega)\omega]$ is the \emph{effective
Boltzmann factor,} with $\beta_{E}(\omega)\equiv\frac{1}{T_{E}(\omega)}$.
Thus, the effective temperature replaces the Hawking temperature in
the equation of the probability of emission \cite{key-17,key-18,key-19,key-20,key-25}.
The effective temperature depends on the energy-frequency of the emitted
radiation and the ratio $\frac{T_{E}(\omega)}{T_{H}}=\frac{2M}{2M-\omega}$
represents the deviation of the BH radiation spectrum from the strictly
thermal feature \cite{key-17,key-18,key-19,key-20,key-25}. It is
better to clarify the definition of effective temperature \cite{key-42}
that we introduced in BH physics in \cite{key-17,key-18}. The probability
of emission of Hawking quanta found by Parikh and Wilczek, i.e. eq.
(\ref{eq: Parikh Correction}), shows that the BH does NOT emit like
a perfect black body, i.e. it has not a strictly thermal behavior.
On the other hand, the temperature in Bose-Einstein and Fermi-Dirac
distributions is a perfect black body temperature. Thus, when we have
deviations from the strictly thermal behavior, i.e. from the perfect
black body, one expects also deviations from Bose-Einstein and Fermi-Dirac
distributions. How can one attack this problem? By analogy with other
various fields of Science, also beyond BHs, for example the case of
planets and stars. One defines the effective temperature of a body
such as a star or planet as the temperature of a black body that would
emit the same total amount of electromagnetic radiation \cite{key-41}.
The importance of the effective temperature in a star is stressed
by the issue that the effective temperature and the bolometric luminosity
are the two fundamental physical parameters needed to place a star
on the Hertzsprung\textendash{}Russell diagram. Both effective temperature
and bolometric luminosity actually depend on the chemical composition
of a star, see again \cite{key-41}.

On the other hand, one recalls that the definition of temperature
in Bose-Einstein or Fermi-Dirac distribution comes from the coefficient
of $\omega$ in the exponential and that is itself ``independent''
of frequency \cite{key-42}. But the key point here is that we have
a deviation from the perfect thermal state and, in turn, we expect
deviations from the exact Bose-Einstein and Fermi-Dirac distributions.
We also stress that the tunnelling is a \emph{discrete} instead of
\emph{continuous} process as two different \emph{countable} BH physical
states have to be considered, the first before the emission of the
particle and the latter after the emission of the particle. The emission
of the particle is, in turn, interpreted like a \emph{quantum} \emph{transition}
of frequency $\omega$ between the two different discrete states.
In fact, the tunnelling mechanism works considering a trajectory in
imaginary or complex time which joins two separated classical turning
points\cite{key-8,key-9}. As a consequence the radiation spectrum
is also discrete. This important issue needs to be clarified in a
better way. At a well fixed Hawking temperature the statistical probability
distribution (\ref{eq: Parikh Correction}) is a continuous function.
But the Hawking temperature in (\ref{eq: Parikh Correction}) varies
in time with a character which is \emph{discrete} because the forbidden
region traversed by the emitting particle has a \emph{finite} size
\cite{key-8,key-9}. If one considers a strictly thermal approximation,
the turning points have zero separation and it is not clear what joining
trajectory has to be considered because there is not barrier \cite{key-8}.
One solves the problem if argues that it is the forbidden finite region
from $r_{initial}=2M\:$ to $r_{final}=2(M\lyxmathsym{\textminus}\omega)\:$
that the tunnelling particle traverses which works like barrier \cite{key-8}.
In other words, the intriguing explanation is that it is the particle
itself which generates a tunnel through the horizon \cite{key-8}.
In this way, one obtains the effective temperature also in the deviation
from the exact Bose-Einstein or Fermi-Dirac distribution. In fact,
by using the definition (\ref{eq: Corda Temperature}) one can easily
rewrite eq. (\ref{eq: final distributions}) as 
\begin{equation}
<n>_{boson}=\frac{1}{\exp\left(\frac{\omega}{T_{E}(\omega)}\right)-1},\;\;<n>_{fermion}=\frac{1}{\exp\left(\frac{\omega}{T_{E}(\omega)}\right)+1},\label{eq: final distributions effettive}
\end{equation}
In other words, one takes an intermediate value between the two subsequent
values of the Hawking temperature, i.e. the effective temperature.
Thus, the introduction of the effective temperature does not degrade
the importance of the Hawking temperature. In fact, as the Hawking
temperature changes with a discrete behavior in time, in a certain
sense the effective temperature represents the value of the Hawking
temperature \emph{during} the emission of the particle. Hence, the
effective temperature takes into account the non-strictly thermal
character of the radiation spectrum and the non-strictly continuous
character of subsequent emissions of Hawking quanta.

\noindent The introduction of the effective temperature permits the
introduction of other \emph{effective quantities}. Considering the
initial BH mass \emph{before} the emission, $M$, and the final BH
mass \emph{after} the emission, $M-\omega$, one introduces the \emph{BH}
\emph{effective mass }and the \emph{BH effective horizon} 
\begin{equation}
M_{E}\equiv M-\frac{\omega}{2},\mbox{ }r_{E}\equiv2M_{E}\label{eq: effective quantities}
\end{equation}

\noindent \emph{during} the BH contraction, i.e. \emph{during} the
emission of the particle \cite{key-17,key-18,key-19,key-20,key-25}.
Such effective quantities are average quantities \cite{key-17,key-18,key-19,key-20,key-25}.
In fact, \emph{$r_{E}$ }is the average of the initial and final horizons
while \emph{$M_{E}$ }is the average of the initial and final masses
\cite{key-17,key-18,key-19,key-20,key-25}. The effective temperature
\emph{$T_{E}\:$ }is the inverse of the average value of the inverses
of the initial and final Hawking temperatures (\emph{before} the emission
$T_{H\mbox{ initial}}=\frac{1}{8\pi M}$, \emph{after} the emission
$T_{H\mbox{ final}}=\frac{1}{8\pi(M-\omega)}$)\cite{key-17,key-18,key-19,key-20,key-25}.
We have also recently shown \cite{key-25} that one can use Hawking's
periodicity argument \cite{key-28,key-29,key-30} to obtain the \emph{effective
Schwarzschild line element }\cite{key-25} 
\begin{equation}
ds_{E}^{2}\equiv-(1-\frac{2M_{E}}{r})dt^{2}+\frac{dr^{2}}{1-\frac{2M_{E}}{r}}+r^{2}(\sin^{2}\theta d\varphi^{2}+d\theta^{2}),\label{eq: Hilbert effective}
\end{equation}
which takes into account the BH \emph{dynamical} geometry during the
emission of the particle. 

\noindent The introduction of $T_{E}(\omega)$ can be applied to the
analysis of the spectrum of BH QNMs \cite{key-17,key-18,key-19,key-20}.
In fact, another key point is that the equation of the BH QNMs frequencies
in \cite{key-21,key-23,key-24} is an approximation as it has been
derived with the assumption that the BH radiation spectrum is strictly
thermal. To take into due account the deviation from the thermal spectrum
one has to replace the Hawking temperature $T_{H}$ with the effective
temperature $T_{E}\:$ in the equation of the BH QNMs frequencies
\cite{key-17,key-18,key-19,key-20}. For large values of the principal
quantum number $n,$ i.e, for excited BHs, and independently of the
angular momentum quantum number, the expression for the quasi-normal
frequencies of the Schwarzschild BH, which takes into account the
non-strictly thermal behavior of the radiation spectrum is \cite{key-17,key-18,key-19,key-20}

\noindent 
\begin{equation}
\begin{array}{c}
\omega_{n}=a+ib+2\pi in\times T_{E}(|\omega_{n}|)\\
\\
\backsimeq2\pi in\times T_{E}(|\omega_{n}|)=\frac{in}{4M-2|\omega_{n}|},
\end{array}\label{eq: quasinormal modes corrected}
\end{equation}
where $a$ and $b$ are real numbers with $a=(\ln3)\times T_{E}(|\omega_{n}|),\; b=\pi\times T_{E}(|\omega_{n}|)$
for $j=0,2$ (scalar and gravitational perturbations), $a=0,\; b=0$
for $j=1$ (vector perturbations) and $a,b\ll|2\pi inT_{E}(|\omega_{n}|)|$.
In complete agreement with Bohr's correspondence principle, it is
trivial to adapt the analysis in \cite{key-21} in the sense of the
Appendix of \cite{key-19} and, in turn, to show that the behavior
(\ref{eq: quasinormal modes corrected}) holds if $j$ is a half-integer
too. A fundamental consequence of eq. (\ref{eq: quasinormal modes corrected})
is that the quantum of area obtained from the asymptotics $|\omega_{n}|$
is an intrinsic property of Schwarzschild BHs because for large $n$
the leading asymptotic behavior of $|\omega_{n}|$ is given by the
leading term in the imaginary part of the complex frequencies, and
it does not depend on the spin content of the perturbation \cite{key-17,key-18,key-19,key-20,key-24}.
An intuitive derivation of eq. (\ref{eq: quasinormal modes corrected})
can be found in \cite{key-17,key-18}. We rigorously derived such
an equation in the Appendix of \cite{key-19}. Eq. (\ref{eq: quasinormal modes corrected})
has the following elegant interpretation \cite{key-17}. The quasi-normal
frequencies determine the position of poles of a Green's function
on the given background, and the Euclidean BH solution converges to
a \emph{non-strictly} thermal circle at infinity with the inverse
temperature $\beta_{E}(\omega_{n})=\frac{1}{T_{E}(|\omega_{n}|)}$
\cite{key-17}. Thus, the spacing of the poles in eq. (\ref{eq: quasinormal modes corrected})
coincides with the spacing $2\pi iT_{E}(|\omega_{n}|)=2\pi iT_{H}(\frac{2M}{2M-|\omega_{n}|}),$
expected for a \emph{non-strictly} thermal Green's function \cite{key-17}.

\section{\noindent Bohr-like model and time dependent Schrödinger equation }

\noindent We found the physical solution for the absolute values of
the frequencies (\ref{eq: quasinormal modes corrected}) in \cite{key-17,key-18}.
One gets

\noindent 
\begin{equation}
E_{n}\equiv|\omega_{n}|=M-\sqrt{M^{2}-\frac{n}{2}}.\label{eq: radice fisica}
\end{equation}
$E_{n}\:$ is interpreted like the total energy emitted by the BH
at that time, i.e. when the BH is excited at a level $n$ \cite{key-17,key-18,key-19,key-20}.
Considering an emission from the ground state to a state with large
$n\:$ and using eq. (\ref{eq: radice fisica}), the BH mass changes
from $M\:$ to 

\begin{equation}
M_{n}\equiv M-E_{n}=\sqrt{M^{2}-\frac{n}{2}}.\label{eq: me-1}
\end{equation}
In the transition from the state with $n$ to a state with $m>n$
the BH mass changes again from $M_{n}\:$ to

\begin{equation}
\begin{array}{c}
M_{m}\equiv M_{n}-\Delta E_{n\rightarrow m}=M-E_{m}\\
=\sqrt{M^{2}-\frac{m}{2}},
\end{array}\label{eq: me}
\end{equation}
where $\Delta E_{n\rightarrow m}\equiv E_{m}-E_{n}=M_{n}-M_{m}$ is
the jump between the two levels due to the emission of a particle
having frequency $\omega_{n,m}=\Delta E_{n\rightarrow m}$. The BH
model that we analysed in \cite{key-17,key-18,key-19} is somewhat
similar to the semi-classical Bohr model of the structure of a hydrogen
atom \cite{key-31,key-32,key-33}. In our BH model \cite{key-17,key-18,key-19},
during a quantum jump a discrete amount of energy is indeed radiated
and, for large values of the principal quantum number $n,$ the analysis
becomes independent of the other quantum numbers. In a certain sense,
QNMs represent the \textquotedbl{}electron\textquotedbl{} which jumps
from a level to another one and the absolute values of the QNMs frequencies
represent the energy \textquotedbl{}shells\textquotedbl{} \cite{key-17,key-18,key-19}.
In Bohr model \cite{key-31,key-32,key-33} electrons can only gain
and lose energy by jumping from one allowed energy shell to another,
absorbing or emitting radiation with an energy difference of the levels
according to the Planck relation $E=hf$, where $\: h\:$ is the Planck
constant and $f\:$ the transition frequency. In our BH model \cite{key-17,key-18,key-19},
QNMs can only gain and lose energy by jumping from one allowed energy
shell to another, absorbing or emitting radiation (emitted radiation
is given by Hawking quanta) with an energy difference of the levels
according to equations which are in full agreement with previous literature
of BH thermodynamics, like references \cite{key-24,key-34,key-35}.
More, the BH model in \cite{key-17,key-18,key-19} is also in agreement
with the famous result of Bekenstein on the \emph{area quantization}
\cite{key-37}. In fact, we found an area quantum arising from a jump
among two neighbouring quantum levels $n-1$ and $n$ having a value
$|\triangle A_{n}|=|\triangle A_{n-1}|\simeq8\pi,$ see eq. (37) in
\cite{key-19}, which is totally consistent with Bekenstein's result
\cite{key-37}. The similarity is completed if one note that the interpretation
of eq. (\ref{eq: radice fisica}) is of a particle, the ``electron'',
quantized with anti-periodic boundary conditions on a circle of length
\cite{key-17} 
\begin{equation}
L=\frac{1}{T_{E}(E_{n})}=4\pi\left(M+\sqrt{M^{2}-\frac{n}{2}}\right),\label{eq: lunghezza cerchio}
\end{equation}
which is the analogous of the electron travelling in circular orbits
around the hydrogen nucleus, similar in structure to the solar system,
of Bohr model \cite{key-31,key-32,key-33}. Clearly, all these similarities
with the Bohr semi-classical model of the hydrogen atom and all these
consistences with well known results in the literature of BHs, starting
by the universal Bekenstein's result, \emph{cannot} be coincidences,
but are confirmations of the correctness of the analysis in \cite{key-17,key-18,key-19}
instead.

On the other hand, Bohr model is an approximated model of the hydrogen
atom with respect to the valence shell atom model of full quantum
mechanics. In the same way, our BH model should be an approximated
model of the quantum BH with respect to the definitive, but at the
present time unknown, model of full quantum gravity theory. 

Now, we show that the energy emitted in an arbitrary transition $n\rightarrow m$,
with $m>n$, is proportional to the effective temperature $\left[T_{E}\right]_{n\rightarrow m}$
associated to the transition. Setting 
\begin{equation}
\Delta E_{n\rightarrow m}\equiv E_{m}-E_{n}=M_{n}-M_{m}=K\left[T_{E}\right]_{n\rightarrow m},\label{eq: differenza radici fisiche}
\end{equation}
where $M_{n}$ and $M_{m}$ are given by eqs. (\ref{eq: me-1}) and
(\ref{eq: me}), let us see if there are values of the constant $K$
for which eq. (\ref{eq: differenza radici fisiche}) is satisfied.
We recall that 

\begin{equation}
\left[T_{E}\right]_{n\rightarrow m}=\frac{1}{4\pi\left(M_{n}+M_{m}\right)},\label{eq: temperatura efficace di transizione}
\end{equation}
because the effective temperature is the inverse of the average value
of the inverses of the initial and final Hawking temperatures \cite{key-17,key-18,key-19,key-20,key-25}.
Thus, eq. (\ref{eq: differenza radici fisiche}) can be rewritten
as 

\begin{equation}
\Delta E_{n\rightarrow m}=M_{n}^{2}-M_{m}^{2}=\frac{K}{4\pi}.\label{eq: differenza radici fisiche 2}
\end{equation}
By using eqs. (\ref{eq: me-1}) and (\ref{eq: me}), for large $m$
and $n$ eq. (\ref{eq: differenza radici fisiche 2}) becomes 

\begin{equation}
\frac{1}{2}\left(m-n\right)=\frac{K}{4\pi},\label{eq: K solved}
\end{equation}
which implies that eq. (\ref{eq: differenza radici fisiche}) is satisfied
for $K=2\pi\left(m-n\right).$ Hence, one finds 
\begin{equation}
\Delta E_{n\rightarrow m}=2\pi\left(m-n\right)\left[T_{E}(\omega_{n,m})\right]_{n\rightarrow m}.\label{eq: differenza radici fisiche finale}
\end{equation}
Using eq. (\ref{eq: Corda Probability}), the probability of emission
between the two levels $n$ and $m$ can be written in the intriguing
form 
\begin{equation}
\Gamma_{n\rightarrow m}=\alpha\exp-\left\{ \frac{\Delta E_{n\rightarrow m}}{\left[T_{E}(\omega)\right]_{n\rightarrow m}}\right\} =\alpha\exp\left[-2\pi\left(m-n\right)\right],\label{eq: Corda Probability Intriguing}
\end{equation}
with $\alpha\sim1.$ Thus, the probability of emission between two
arbitrary levels characterized by the two ``overtone'' quantum numbers
$n$ and $m$ scales like $\exp\left[-2\pi\left(m-n\right)\right].$
In particular, for $m=n+1$ the probability of emission has its maximum
value $\sim\exp(-2\pi)$, i.e. the probability is maximum for two
adjacent levels, as one intuitively expects. 

From the quantum mechanical point of view, one physically interprets
Hawking radiation like energies of quantum jumps among the unperturbed
levels (\ref{eq: radice fisica}) \cite{key-17,key-18,key-19,key-20}.
In quantum mechanics, time evolution of perturbations can be described
by an operator \cite{key-26}

\emph{
\begin{equation}
U(t)=\begin{array}{c}
W(t)\;\;\; for\;0\leq t\leq\tau\\
0\;\;\; for\; t<0\; and\; t>\tau.
\end{array}\label{eq: perturbazione}
\end{equation}
}Then, the complete (time dependent) Hamiltonian is described by the
operator \cite{key-26}

\begin{equation}
H(x,t)\equiv V(x)+U(t),\label{eq: Hamiltoniana completa}
\end{equation}
where $V(x)$ is the \emph{effective Regge-Wheeler} potential of the
time independent Schröedinger-like equation which governs QNMs, see
\cite{key-17,key-18,key-19} for details. Thus, for a wave function
$\psi(x,t),$ one can write the correspondent \emph{time dependent
Schroedinger equation }for the system \cite{key-26}

\begin{equation}
i\frac{d|\psi(x,t)>}{dt}=\left[V(x)+U(t)\right]|\psi(x,t)>=H(x,t)|\psi(x,t)>.\label{eq: Schroedinger equation}
\end{equation}
The\emph{ }state\emph{ }which satisfies eq. (\ref{eq: Schroedinger equation})
is \cite{key-26}

\begin{equation}
|\psi(x,t)>=\sum_{m}a_{m}(t)\exp\left(-i\omega_{m}t\right)|\varphi_{m}(x)>,\label{eq: Schroedinger wave-function}
\end{equation}
where the $\varphi_{m}(x)$ are the eigenfunctions of the time independent
Schröedinger-like equation in {[}\cite{key-17,key-18,key-19}, and
the $\omega_{m}$ are the correspondent eigenvalues. One considers
Dirac delta perturbations \cite{key-17,key-18,key-19,key-20,key-24}
which represent subsequent absorptions of particles having negative
energies which are associated to emissions of Hawking quanta in the
mechanism of particle pair creation. Thus, in the basis $|\varphi_{m}(x)>$,
the matrix elements of $W(t)$ can be written as

\begin{equation}
W_{ij}(t)\equiv A_{ij}\delta(t),\label{eq: a delta}
\end{equation}
where $W_{ij}(t)=<\varphi_{i}(x)|W(t)|\varphi_{j}(x)>$ \cite{key-26}
and the $A_{ij}$ are real. In order to solve the complete quantum
mechanical problem described by the operator (\ref{eq: Hamiltoniana completa})
one needs to know the probability amplitudes $a_{m}(t)$ due to the
application of the perturbation described by the time dependent operator
(\ref{eq: perturbazione}) \cite{key-26}, which represents the perturbation
associated to the emission of an Hawking quantum. For $t<0,$ i.e.
before the perturbation operator (\ref{eq: perturbazione}) starts
to work, the system is in a stationary state $|\varphi_{n}(t,x)>,$
at the quantum level $n,$ with energy $E_{n}=|\omega_{n}|$ given
by eq. (\ref{eq: radice fisica}). Therefore, in eq. (\ref{eq: Schroedinger wave-function})
only the term

\begin{equation}
|\psi_{n}(x,t)>=\exp\left(-i\omega_{n}t\right)|\varphi_{n}(x)>,\label{eq: Schroedinger wave-function in.}
\end{equation}
is not null for $t<0.$ This implies $a_{m}(t)=\delta_{mn}\:\:$for
$\: t<0.$ When the perturbation operator (\ref{eq: perturbazione})
stops to work, i.e. after the emission, for $t>\tau$ the probability
amplitudes $a_{m}(t)$ return to be time independent, having the value
$a_{n\rightarrow m}(\tau)$ \cite{key-26}. In other words, for $t>\tau\:$
the system is described by the\emph{ wave function $\psi_{final}(x,t)$
}which corresponds to the state

\begin{equation}
|\psi_{final}(x,t)>=\sum_{m=n}^{m_{max}}a_{n\rightarrow m}(\tau)\exp\left(-i\omega_{m}t\right)|\varphi_{m}(x)>.\label{eq: Schroedinger wave-function fin.}
\end{equation}
Thus, the probability to find the system in an eigenstate having energy
$E_{m}=|\omega_{m}|$ is given by \cite{key-26}

\begin{equation}
\Gamma_{n\rightarrow m}(\tau)=|a_{n\rightarrow m}(\tau)|^{2}.\label{eq: ampiezza e probability}
\end{equation}
By using a standard analysis, one obtains the following differential
equation from eq. (\ref{eq: Schroedinger wave-function fin.}) \cite{key-26}

\begin{equation}
i\frac{d}{dt}a_{n\rightarrow m}(t)=\sum_{l=m}^{m_{max}}W_{ml}a_{n\rightarrow l}(t)\exp\left[i\left(\Delta E_{l\rightarrow m}\right)t\right].\label{eq: systema differenziale}
\end{equation}
To first order in $U(t)$, by using the Dyson series, one gets the
solution \cite{key-26}

\begin{equation}
a_{n\rightarrow m}=-i\int_{0}^{t}\left\{ W_{mn}(t')\exp\left[i\left(\Delta E_{n\rightarrow m}\right)t'\right]\right\} dt'.\label{eq: solution}
\end{equation}
By inserting (\ref{eq: a delta}) in (\ref{eq: solution}) one obtains
\begin{equation}
a_{n\rightarrow m}=iA_{mn}\int_{0}^{t}\left\{ \delta(t')\exp\left[i\left(\Delta E_{n\rightarrow m}\right)t'\right]\right\} dt'=\frac{i}{2}A_{mn}\label{eq: solution 2}
\end{equation}
Combining this equation with eqs. (\ref{eq: Corda Probability Intriguing})
and (\ref{eq: ampiezza e probability}) one gets 

\begin{equation}
\begin{array}{c}
\alpha\exp\left[-2\pi\left(m-n\right)\right]=\frac{1}{4}A_{mn}^{2}\\
\\
A_{mn}=2\sqrt{\alpha}\exp\left[-\pi\left(m-n\right)\right]\\
\\
a_{n\rightarrow m}=-i\sqrt{\alpha}\exp\left[-\pi\left(m-n\right)\right].
\end{array}\label{eq: uguale}
\end{equation}
As $\sqrt{\alpha}\sim1,$ one gets $A_{mn}\sim10^{-2}$ for $m=n+1$,
i.e. when the probability of emission has its maximum value. This
implies that second order terms in $U(t)$ are $\sim10^{-4},$ i.e.
the approximation is very good. Clearly, for $m>n+1$ the approximation
is better because the $A_{mn}$ are even smaller than $10^{-2}$.
Thus, one can write down the final form of the ket representing the
state as

\begin{equation}
|\psi_{final}(x,t)>=\sum_{m=n}^{m_{max}}-i\sqrt{\alpha}\exp\left[-\pi\left(m-n\right)-i\omega_{m}t\right]|\varphi_{m}(x)>.\label{eq: Schroedinger wave-function finalissima}
\end{equation}
The\emph{ }state (\ref{eq: Schroedinger wave-function finalissima})
represents a \emph{pure final state instead of a mixed final state.}
Hence, the states are written in terms of an \emph{unitary} evolution
matrix instead of a density matrix\emph{ }and\emph{ }this implies
the fundamental conclusion that\emph{ }information is not loss in
BH evaporation. The result agrees with the assumption by 't Hooft
that Schrödinger equations can be used universally for all dynamics
in the universe \cite{key-6}. We also stress that, as the final state
of eq. (\ref{eq: Schroedinger wave-function finalissima}) is due
to potential emissions of Hawking quanta having negative energies
which perturb the BH and ``trigger'' the QNMs corresponding to potential
arbitrary transitions $n\rightarrow m$, with $m>n$, the subsequent
\emph{collapse of the wave function} to a new a stationary state 
\begin{equation}
|\psi_{m}(x,t)>=\exp\left(-i\omega_{m}t\right)|\varphi_{m}(x)>,\label{eq: Schroedinger wave-function out}
\end{equation}
at the quantum level $m,$ implies that the wave function of the particle
having negative energy $-\Delta E_{n\rightarrow m}=\omega_{n}-\omega_{m}$
has been transferred to the QNM and it is given by 
\begin{equation}
|\psi_{-\left(m-n\right)}(x,t)>\equiv-\exp\left[i(\omega_{m}-\omega_{n})t\right]\left[|\varphi_{m}(x)>-|\varphi_{n}(x)>\right].\label{eq: funzione onda particella emessa}
\end{equation}
This wave function results entangled with the wave function of the
particle with positive energy which propagates towards infinity in
the mechanism of particle creation by BHs. We will see in the following
Section that this key point solves the entanglement problem connected
with the information paradox.

The analysis in this work is strictly correct only for $n\gg1,$ i.e.
only for excited BHs. This is the reason because we assumed an emission
from the ground state to a state with large $n\:$ in the discussion.
On the other hand, a state with large $n\:$ is always reached at
late times, maybe not through a sole emission from the ground state,
but, indeed, through various subsequent emissions of Hawking quanta. 

For the sake of completeness, it is helpful to discuss on the exact
nature of time as it is defined in the Schrodinger equation and the
covariant properties of that equation \cite{key-43}. Following \cite{key-26},
we recall that time must be considered as a parameter in quantum mechanics
instead of a quantum mechanical operator. In particular, time is \emph{not}
a quantum observable \cite{key-26}. In other words, in quantum mechanics
it is \emph{not} possible to discuss on a ``time operator'' in the
same way that we do, for example for the ``position operator'' and
for the ``momentum operator'' \cite{key-26}. In fact, within the
framework of quantum mechanics there is no room for a symmetric analysis
for both time and position, even if, from an historical point of view,
quantum mechanics has been developed by De Broglie and Schroedinger
following the idea of a covariant analogy between time and energy
on one hand and position and momentum on the other hand \cite{key-26}.
This discussion works from the quantum mechanical point of view of
the analysis. But, in the current analysis, we have to discuss about
time also from the point of view of general relativity. As we discussed
Hawking radiation as tunnelling in the framework of the analysis in
\cite{key-8,key-9} we recall that in such works the Painle\'{v} and
Gullstrand coordinates for the Schwarzschild geometry have been used.
On the other hand, the radial an time coordinates are the same in
both the Painle\'{v} and Gullstrand and Schwarzschild line elements.
Thus, we conclude that the time in the operator for the time evolution
of eq. (\ref{eq: perturbazione}) and in the subsequent equations
is the Schwarzschild time.

Concerning the covariant properties of the Schrodinger equation, we
recall that, in general, quantum mechanics has to be applied to systems
moving with speeds that are not negligible with respect to the speed
of light. Then, relativistic corrections could be in principle necessary
in various cases. This important issue implies that the laws of quantum
mechanics, and, in turn, the Schrodinger equation, must be re-formuled
in covariant form through the Lorentz transformations. The conseguence
will be the apparence of news important quantum properties like the
spin and the spin-orbit interaction. This is not the case of the analysis
in this paper because we use Bohr's Correspondence Principle which
enables an accurate semi-classical analysis for large values of the
principal quantum number $n.$

\section{Solution to the entanglement problem}

One could claim that, although the above analysis provides a natural
model of Hawking radiation, it makes no reference to the BH spacetime,
where information is conserved. In fact, some authors claim that the
challenge in addressing information loss is to reconcile models of
Hawking radiation with the spacetime structure, where the information
falling into the singularity is causally separated from the outgoing
Hawking radiation, see \cite{key-13} for example. In any case, these
criticisms do not work for the analysis in this work. In fact, in
the above analysis there is a subtle connection between Hawking radiation
and the BH spacetime, where information is conserved. The key point
of this approach to the Hawking information problem concerns the entanglement
structure of the wave function associated to the particle pair creation
\cite{key-13}. In other terms, in order to solve the paradox, one
needs to know the part of the wave function in the interior of the
horizon \cite{key-13}, i.e. the part of the wave function associated
to the particle having negative energy (interior, infalling modes).
This is exactly the part of the wave function which in the Hawking
computation gets entangled with the part of the wave function outside,
i.e. the part of the wave function associated to the particle having
positive energy which escapes from the BH \cite{key-13}. If one ignores
this interior part of the wave function, one misses the entanglement
completely, and thus fails to understand the paradox \cite{key-13}.
But in the correspondence between Hawking radiation and BH QNMs the
particle having negative energy which falls into the singularity transfers
its part of the wave function and, in turn, the information encoded
in such a part of the wave function, to the QNM. Hence, the emitted
radiation results to be entangled with BH QNMs, which are the oscillations
of the BH horizon. This fundamental point is exactly the subtle connection
between the emitted radiation and the interior BH spacetime that one
needs to find. In other words, although we do not know what happens
in the interior spacetime structure, we know that the response of
such a structure to the absorption of an interior, infalling mode
is to add the frequency of that interior, infalling mode (and, in
turn, of its wave function) to the QNM corresponding to the energy
level $E_{n}$, in order to permit it to jump to the energy level
$E_{m}$. In that way, the interior part of the wave function is now
``within'' the QNM corresponding to the quantum level $E_{m}$,
which is, in turn, entangled with all the particles emitted at that
time. Let us see this issue in detail. Once again, we stress that
the correspondence between emitted radiation and proper oscillation
of the emitting body is a fundamental behavior of every radiation
process in Nature, and this is the key point which permits to solve
the entanglement problem. One describes the mechanism of particles
creation by BHs \cite{key-1}, as tunnelling arising from vacuum fluctuations
near the BH horizon \cite{key-8,key-9,key-25}. If a virtual particle
pair is created just inside the horizon, the virtual particle with
positive energy can tunnel out. Then, it materializes outside the
BH as a real particle. In the same way, if one considers a virtual
particle pair created just outside the horizon, the particle with
negative energy can tunnel inwards. In both of the situations, the
particle with negative energy is absorbed by the BH. Again, let us
assume a first emission from the BH ground state to a state with large
$n,$ say $n=n_{1}\gg1.$ The absorbed particle having negative energy
$-|\omega_{n_{1}}|$ generates a QNM corresponding to an energy-level
of emitted energies $E_{n_{1}}=|\omega_{n_{1}}|$ and the BH mass
changes from $M\:$ to 

\begin{equation}
M_{n_{1}}\equiv M-E_{n_{1}}=\sqrt{M^{2}-\frac{n_{1}}{2}}.\label{eq: m1}
\end{equation}
In other words, the energy of the first absorbed particle having negative
energy is transferred, together with its part of the wave function,
to the QNM which is now entangled with the emitted particle having
positive energy. By using eq. (\ref{eq: funzione onda particella emessa})
and setting $n=0$ and $m=n_{1}$ one finds that the part of the wave
function in the interior of the horizon, i.e. the part of the wave
function associated to the particle having negative energy (infalling
mode) which has been transferred to the QNM is 
\begin{equation}
|\psi_{-n_{1}}(x,t)>=-\exp\left(i\omega_{n_{1}}t\right)|\varphi_{n_{1}}(x)>.\label{eq: da zero ad n1}
\end{equation}
Now, let us consider a second emission, which corresponds to the transition
from the state with $n=n_{1}$ to a state with, say, $n=n_{2}>n_{1}$.
The BH mass changes from $M_{n_{1}}\:$ to

\begin{equation}
\begin{array}{c}
M_{n_{2}}\equiv M_{n_{1}}-\Delta E_{n_{1}\rightarrow n_{2}}=M-E_{n_{2}}\\
=\sqrt{M^{2}-\frac{n_{2}}{2}},
\end{array}\label{eq: m2}
\end{equation}
where $\Delta E_{n_{1}\rightarrow n_{2}}\equiv E_{n_{2}}-E_{n_{1}}=M_{n_{1}}-M_{n_{2}}$
is the jump between the two levels. The energy of the second absorbed
particle having negative energy is transferred, together with its
part of the wave function, again to the QNM, which now corresponds
to an increased level of energy $E_{n_{2}}=|\omega_{n_{12}}|$ and
is now entangled with both the two emitted particles having positive
energy. By using again eq. (\ref{eq: funzione onda particella emessa})
and setting $n=n_{1}$ and $m=n_{2}$, one finds that the part of
the wave function of the second infalling mode which has been transferred
to the QNM is 
\begin{equation}
|\psi_{-\left(n_{2}-n_{1}\right)}(x,t)>=-\exp\left[i(\omega_{n_{2}}-\omega_{n_{1}})t\right]\left[|\varphi_{n_{2}}(x)>-|\varphi_{n_{1}}(x)>\right].\label{eq: da n1 ad n2}
\end{equation}
Let us consider a third emission, which corresponds to the transition
from the state with $n=n_{2}$ to a state with, say, $n=n_{3}>n_{2}$.
The BH mass changes from $M_{n_{2}}\:$ to

\begin{equation}
\begin{array}{c}
M_{n_{3}}\equiv M_{n_{2}}-\Delta E_{n_{2}\rightarrow n_{3}}=M-E_{n_{3}}\\
=\sqrt{M^{2}-\frac{n_{3}}{2}},
\end{array}\label{eq: m3}
\end{equation}
where $\Delta E_{n_{2}\rightarrow n_{3}}\equiv E_{n_{3}}-E_{n_{2}}=M_{n_{2}}-M_{n_{3}}$
is the jump between the two levels. Again, the energy of the third
absorbed particle having negative energy is transferred, together
with its part of the wave function, to the QNM corresponding now to
a further increased energy level $E_{n_{13}}=|\omega_{n_{3}}|$ and
being entangled with the three emitted particles which have positive
energy. Now, eq. (\ref{eq: funzione onda particella emessa}) with
$n=n_{2}$ and $m=n_{3}$ gives the part of the wave function of the
third infalling mode which has been transferred to the QNM as 
\begin{equation}
|\psi_{-\left(n_{3}-n_{2}\right)}(x,t)>=-\exp\left[i(\omega_{n_{3}}-\omega_{n_{2}})t\right]\left[|\varphi_{n_{3}}(x)>-|\varphi_{n_{12}}(x)>\right].\label{eq: da n2 ad n3}
\end{equation}
The process will continue again, and again, and again... till the
\emph{Planck distance} and the \emph{Planck mass} are approached by
the evaporating BH. At that point, the Generalized Uncertainty Principle
prevents the total BH evaporation in exactly the same way that the
Uncertainty Principle prevents the hydrogen atom from total collapse
\cite{key-27} and one needs a full theory of quantum gravity for
the further evolution. 

In any case, we stress again that the energy $E_{n}\:$ of the generic
QNM having quantum ``overtone'' number $n$ is interpreted like
the total energy emitted by the BH at that time, i.e. when the BH
is excited at a level $n$ \cite{key-17,key-18,key-19,key-20}. This
implies that such a QNM is entangled with all the Hawking quanta emitted
at that time.

Thus, all the quantum physical information falling into the singularity
is not causally separated from the outgoing Hawking radiation, but
it is instead recovered and codified in eq. (\ref{eq: Schroedinger wave-function finalissima})
which leads the time evolution of the correspondence between Hawking
radiation and BH QNMs. In other words, in our approach the ``smoothness
of the horizon'' is achieved by the issue that the horizon is oscillating
through QNMs and all the emitted particles are entangled with such
oscillations. As the solution to the information problem should be
to find a physical effect that one might have have missed \cite{key-13},
here we have shown that the natural correspondence between Hawking
radiation and BH QNMs \emph{is exactly that missed physical effect}.

\section{Conclusion remarks}

Through an analysis of BH QNMs in terms of an unitary evolution, governed
by a time dependent Schroedinger equation of a Bohr-like model for
BHs as ``hydrogen atoms, '' we falsified Hawking's claim on the
information loss in BH evaporation. We stress that it is an intuitive
but general conviction that BHs result in highly excited states representing
both the ``hydrogen atom'' and the ``quasi-thermal emission''
in quantum gravity. Here we have shown that such an intuitive picture
is more than a picture, showing that a model of quantum BH somewhat
similar to the historical semi-classical model of the structure of
a hydrogen atom introduced by Bohr in 1913 \cite{key-31,key-32,key-33}
has a time evolution obeying a time dependent Schrödinger equation,
in perfect agreement with quantum mechanics. Clearly, this cannot
be a coincidence. In the same way, they cannot be coincidences the
consistences with various papers in the literature of BH thermodynamics,
see for example \cite{key-24,key-34,key-35} and the consistence with
the famous result of Bekenstein on the area quantization \cite{key-37}. 

In the final Section of this paper we have also show that the present
approach permits to solve the entanglement problem connected with
the information paradox. 

We also recall that consistence between the time evolution of our
Bohr-like model and a recent approach to solve the BH information
paradox \cite{key-10,key-11,key-12,key-14} has been recently highlighted
in \cite{key-40}.

\section{Acknowledgements }

The Scuola Superiore Internazionale di Studi Universitari e Ricerca
``Santa Rita'' has to be thanked for supporting this paper. I thank
two unknown referees for useful comments.


\begin{thebibliography}{10}
\bibitem{key-1}S. W. Hawking, Commun. Math. Phys. 43, 199 (1975).

\bibitem{key-2}S. W. Hawking, Phys. Rev. D 14, 2460 (1976). 

\bibitem{key-3}S. W. Hawking, Phys. Rev. D 72, 084013 (2005).

\bibitem{key-4}L. Susskind, \emph{The Black Hole War: My Battle with
Stephen Hawking to Make the World Safe for Quantum Mechanics}, Little,
Brown and Company (2008).

\bibitem{key-5}A. Ananthaswamy, \emph{Black Holes: Paradox Regained},
http://www.fqxi.org/

\bibitem{key-6}G. 't Hooft, Int. J. Mod. Phys. A 11, 4623-4688 (1996). 

\bibitem{key-7}Samir D. Mathur, arXiv:1201.2079. (Expanded version
of) proceedings for Lepton-Photon 2011. 

\bibitem{key-8}M. K. Parikh, Gen. Rel. Grav. 36, 2419 (2004, First
Award in the Gravity Research Foundation Essay Competition).

\bibitem{key-9}M. K. Parikh and F. Wilczek, Phys. Rev. Lett. 85,
5042 (2000).

\bibitem{key-10}B. Zhang, Q.-Y. Cai, L. You,, and M. S. Zhan, Phys.
Lett. B 675, 98 (2009). 

\bibitem{key-11}B. Zhang, Q.-Y. Cai, M. S. Zhan, and L. You, Ann.
Phys. 326, 350 (2011). 

\bibitem{key-12}B. Zhang, Q.-Y. Cai, M. S. Zhan, and L. You, EPL
94, 20002 (2011). 

\bibitem{key-13}Samir D. Mathur, arXiv:1108.0302v2 (hep-th). 

\bibitem{key-14}B. Zhang, Q.-Y. Cai, M. S. Zhan, and L. You, arXiv:1210.2048.

\bibitem{key-15}B. Zhang, Q.-Y. Cai, M. S. Zhan, and L. You, D 22,
1341014 (2013, First Award in the Gravity Research Foundation Essay
Competition).

\bibitem{key-16}Samir D. Mathur, Int. Journ. Mod. Phys. D 22, 1341016
(2013, Third Award in the Gravity Research Foundation Essay Competition).

\bibitem{key-17}C. Corda, Int. Journ. Mod. Phys. D 21, 1242023 (2012,
Honorable Mention in the Gravity Research Foundation Competition).

\bibitem{key-18}C. Corda, JHEP 08 (2011) 101. 

\bibitem{key-19}C. Corda, Eur. Phys. J. C 73, 2665 (2013).

\bibitem{key-20}C. Corda, S. H. Hendi, R. Katebi, N. O. Schmidt,
JHEP 06 (2013) 008. 

\bibitem{key-21}L. Motl, Adv. Theor. Math. Phys. 6, 1135 (2003).

\bibitem{key-22}J. York Jr., Phys. Rev. D28, 2929 (1983). 

\bibitem{key-23}S. Hod, Phys. Rev. Lett. 81 4293 (1998); S. Hod,
Gen. Rel. Grav. 31, 1639 (1999, Fifth Award in the Gravity Research
Foundation Competition).

\bibitem{key-24}M. Maggiore, Phys. Rev. Lett. 100, 141301 (2008). 

\bibitem{key-25}C. Corda, Ann. Phys. 337, 49 (2013), definitive version
with corrected typos in arXiv:1305.4529v3.

\bibitem{key-26}J. J. Sakurai, \emph{Modern Quantum Mechanic}s, Pearson
Education (2006).

\bibitem[27]{key-27}R. J. Adler, P. Chen and D. I. Santiago, Gen.
Rel. Grav. 3, 2101-2108 (2001, Third Award in the Gravity Research
Foundation Essay Competition). 

\bibitem[28]{key-28}R. Banerjee and B. R. Majhi, Phys. Rev. D 79,
064024 (2009).

\bibitem[29]{key-29}R. Banerjee and B. R. Majhi, Phys. Lett. B 674,
218 (2009).

\bibitem[30]{key-30}S. W. Hawking, \textquotedblleft{}\emph{The Path
Integral Approach to Quantum Gravity}\textquotedblright{}, in General
Relativity: An Einstein Centenary Survey, eds. S.W. Hawking and W.
Israel, (Cambridge University Press, 1979). 

\bibitem[31]{key-31}N. Bohr, Philos. Mag. 26 , 1 (1913).

\bibitem[32]{key-32}N. Bohr, Philos. Mag. 26 , 476 (1913).

\bibitem[33]{key-33}N. Bohr, Zeits. Phys. 2, 423 (1920). 

\bibitem[34]{key-34}S. Shankaranarayanan, Mod. Phys. Lett. A 23,
1975-1980 (2008). 

\bibitem[35]{key-35}J. Zhang, Phys. Lett. B 668, 353-356 (2008). 

\bibitem[36]{key-36}S. W. Hawking, arXiv:1401.5761 (2014).

\bibitem[37]{key-37}J. D. Bekenstein, Lett. Nuovo Cim. 11, 467 (1974). 

\bibitem[38]{key-38}C. Corda, S. H. Hendi, R. Katebi, N. O. Schmidt,
Adv. High En. Phys. 527874 (2014).

\bibitem[39]{key-39}C. Corda, S. H. Hendi, R. Katebi, N. O. Schmidt,
Adv. High En. Phys. 530547 (2014).

\bibitem[40]{key-40}X.-K. Guo, Q.-Y., Cai, Int. Journ. Theor. Phys.,
53, 2980 (2014). 

\bibitem[41]{key-41}A. E. Roy and D. Clarke, Astronomy: Principles
and Practice, Fourth Edition (2003, PBK).

\bibitem[42]{key-42}Private communication with the Reviewer \#1.

\bibitem[43]{key-43}Private communication with the Reviewer \#2.\end{thebibliography}
\end{document}